\documentclass[lettersize,journal]{IEEEtran}

\IEEEoverridecommandlockouts
\usepackage{cite}
\usepackage{amsmath,amssymb,amsfonts}
\usepackage{amsmath}
\usepackage{graphicx}
\usepackage{textcomp}
\usepackage{xcolor}
\usepackage{booktabs}
\usepackage[colorlinks=true,bookmarks=false,citecolor=blue,urlcolor=blue]{hyperref} 
\usepackage{multirow}
\usepackage{tabularx}
\usepackage{algorithm,algpseudocode}
\usepackage{algpseudocode} 
\usepackage{comment}

\begin{document}

\title{A Queuing Envelope Model for Estimating Latency Guarantees in Deterministic Networking Scenarios}

\author{
\IEEEauthorblockN{Nataliia Koneva\IEEEauthorrefmark{1}, Alfonso S\'{a}nchez-Maci\'{a}n\IEEEauthorrefmark{1}, 
Jos\'{e} Alberto Hern\'{a}ndez\IEEEauthorrefmark{1},
Farhad Arpanaei\IEEEauthorrefmark{1},
\'{O}scar Gonz\'{a}lez de Dios\IEEEauthorrefmark{2}
}
\IEEEauthorblockA{\IEEEauthorrefmark{1}Dept. Ing. Telemática, Universidad Carlos III de Madrid, Spain, \IEEEauthorrefmark{2} Telefonica I+D, Spain.
}
}



\maketitle

\begin{abstract}



Accurate estimation of queuing delays is crucial for designing and optimizing communication networks, particularly in the context of Deterministic Networking (DetNet) scenarios. This study investigates the approximation of Internet queuing delays using an M/M/1 envelope model, which provides a simple methodology to find tight upper bounds of real delay percentiles. Real traffic statistics collected at large Internet Exchange Points (like Amsterdam and San Francisco) have been used to fit polynomial regression models for transforming packet queuing delays into the M/M/1 envelope models. We finally propose a methodology for providing delay percentiles in DetNet scenarios where tight latency guarantees need to be assured.


\end{abstract}


\section{Introduction}
\label{Introduction}

Deterministic networking (DetNet) is a new paradigm for network design and management that seeks to ensure predictable and reliable network performance, including bounded latency and reliability~\cite{detnet_5g_application,100Gbps_DetNet}. This approach can be useful for applications that require low latency and high throughput, such as real-time communications and industrial automation~\cite{LARRABEITI2023103220,gonzalo_rtt}. However, deterministic networks are often challenging to design and analyze due to the complexity of the queuing behavior of network devices. In particular, queuing delays can be difficult to model and predict accurately, especially in scenarios where traffic is bursty and the number of independent uncorrelated sources is small~\cite{tail_at_scale}. 

DetNet aims to provision deterministic data paths for real-time applications, ensuring extremely low packet loss rates, low jitter, and bounded end-to-end latency for Quality-of-Service (QoS) assurance. A critical issue in providing deterministic services is allocating the right amount of resources to ensure quality of service (QoS) requirements within a network slice~\cite{deterministic_latency}. The authors in~\cite{detnet_5g_application} discussed the necessity for low and controlled latency in 5G networks to support time-sensitive applications. For instance, in cloud gaming, an end-to-end latency of $50-80~ms$ ensures a smooth experience, while $120~ms$ can significantly impact responsiveness. Thus, DetNet segments should have slotted, scheduled, and synchronous architectures to cover these strict performance requirements. Other scenarios like telemedicine~\cite{telesurgery}, connected vehicles, and Industrial IoT~\cite{Broadband_Technologies} demand top connectivity guarantees in terms of packet loss and latency. This has given rise to defining strategies for Time-Sensitive Networking (TSN) which, in combination with DetNet, can provide near-lossless packet forwarding with deterministic latency in 100~Gb/s scenarios~\cite{100Gbps_DetNet}.






In the past years, network traffic has been revisited and it has been found that traffic volumes approximate the Gaussian distribution~\cite{gaussianity_v1} and~\cite{gaussianity_v2}, and they can be even better characterized by a log-normal distribution, especially in scenarios where high aggregation of multiple independent sources occur~\cite{alasmar_traffic}. In fact, packet inter-arrival times can be approximated by a non-homogeneous Poisson process, and the M/G/1 queuing model can be applied for characterising packet delays~\cite{karagiannis}. However, the M/G/1 model does not have a closed-form expression for the cumulative density function (CDF) of queuing delays, thus making the task of computing delay percentiles and worst-case latency values difficult for network modelers. This work aims to provide a simple queuing delay model that can be used as an upper bound for estimating end-to-end latency bounds in DetNet scenarios.

\section{Background and Methodology}
\label{Methodology}

The simplest delay model for network engineers is the well-known M/M/1 queuing model, where packets are assumed to arrive following a Poisson process with rate $\lambda$ packet/s and packet service times $X$ are exponentially distributed with a mean of $\mu^{-1}$ seconds per packet. In M/M/1 models, the CDF of queuing plus transmission delay $D$ per queue follows an exponential distribution:
\begin{equation}
\label{eq:mm1_cdf}
    F_D (t) = 1 - e^{-\mu (1-\rho)t},\quad t\ge 0
\end{equation}
where the server's load $\rho = \frac{\lambda}{\mu}$ is the ratio between the packet arrival intensity $\lambda$ (in packet/s) and the service rate $\mu$ (in packet/s). Remark that $\rho$ must be smaller than unity to ensure system's stability, and the average packet service time is computed as $E(X)=\frac{1}{\mu} = \frac{8\bar{L}}{C}$, where $\bar{L}$ is the average packet length (in bytes) and $C$ is the server's bit rate (in bps). In the M/M/1 queuing model, there is also an exact formula for obtaining delay percentiles $D_{\text{q}}$, as:
\begin{equation}
    \label{eq:quantile}
    D_{\text{q}} = E(X) \frac{1}{1-\rho}\ln \frac{1}{1-q}
\end{equation}

However, in real network traffic, service times are not exponentially distributed instead, they follow some generic packet size distribution with multiple typical packet lengths. For instance, the Amsterdam Internet Exchange Point (AMS-IXP) shows a packet size distribution with a mean of $\bar{L}=1019.03$ bytes and a standard deviation of $\sigma_L = 1161.66$ bytes (thus the squared coefficient of variation $C_L^2 = \frac{\sigma_L^2}{\bar{L}^2} = 1.30$) \cite{amsix2024}. The San Francisco SFM-IXP shows a mean of 1750.41 bytes and a standard deviation of 2062.69 bytes (thus a squared coefficient of variation of 1.39) \cite{sfmix2024}. These distributions capture the diversity of packet sizes encountered on the Internet.

In such cases, the M/G/1 model needs to be applied but, unfortunately, there is no closed-form solution for the CDF like in the M/M/1 case, only the Pollaczek-Khintchine formula for the average waiting time in queue $W_q$ as:
\begin{equation}\label{eq:mg1} 
E(W_q)_{\text{M/G/1}} = E(X) \frac{\rho}{1-\rho}\frac{1+C_X^2}{2} 
\end{equation}
but does not provide a simple equation for the CDF or ways to obtain theoretical equations for delay percentiles. 

Thus, the goal of this article is to find the closest M/M/1 queuing model (we call it queuing envelope) that acts as an upper bound for real M/G/1 queues with empirically-observed packet size distributions, like those from AMS-IXP and SFM-IXP. To do this, we define Algorithm~\ref{alg:envelope_calculation} which seeks the smallest value of $\rho_{env}$ or $\rho_{M/M/1}$ that acts as an upper bound for a real M/G/1 queue whose load is $\rho_{real}$ or $\rho_{M/G/1}$. In other words, we aim to approximate an M/G/1 queue fed with real network traffic by the closest M/M/1 queuing model whose delay is always above the real one for percentiles above the median (50\%), therefore acting as an upper bound of real delay:

$$\textrm{Find smallest }\rho_{\text{env}}\textrm{ such that }D_{\text{env}}\ge D_{\text{real}}$$
$$\textrm{ for delay percentiles above 50\%}$$
Here, $D_{\text{real}}$ refers to the real delay observed in an M/G/1 queue fed with real network traffic, and $D_{\text{env}}$ refers to the closest upper bound M/M/1 envelope delay found by Algorithm~\ref{alg:envelope_calculation}.

\begin{algorithm}
\caption{Envelope M/M/1 Load Calculation}
\label{alg:envelope_calculation}
\begin{algorithmic}[1]
\State \textbf{Input:} 
\State $mg1\_packets$: simulated packet delays from M/GI/1 system
\State $E_X$: average service time for M/GI/1 system
\State $E_T\_real$: average delay for M/GI/1 system

\State \textbf{Initialize} $percentiles\_seq \leftarrow$ seq(0.5, 0.99, 0.01)  \Comment{from $50\%$ to $99\%$}
\State \textbf{Calculate} $df\_real$: Real M/GI/1 quantiles from $50\%$ to $99\%$ for $mg1\_packets$
\State \textbf{Initialize} $\rho_{env} \leftarrow$ seq(from=0.01, to=0.99, by=0.01) \Comment{Candidate M/M/1 loads: from 0.01 to 0.99 with step 0.01}
\For{each $\rho$ in $\rho_{env}$}
    \State \textbf{Calculate} $df\_env \leftarrow$ qexp($percentiles\_seq$, $\lambda_{rate} =\frac{1-\rho}{E_X}$
    M/M/1 envelope quantiles from $50\%$ to $99\%$
    \If{all($df\_real < df\_env$)}
        \State \textbf{Return} $\rho$  and $\frac{E_X}{1-\rho}$
    \EndIf
\EndFor
\State \textbf{Output:} Envelope M/M/1 load ($\rho_{env}$) and its average delay
\end{algorithmic}
\end{algorithm}

\begin{figure*}[!htbp]
   \centering
    \includegraphics[width=0.48\linewidth]{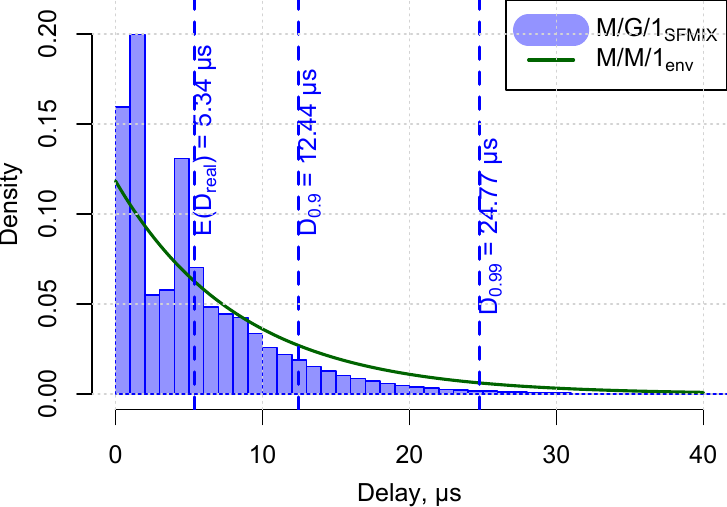}
    \includegraphics[width=0.48\linewidth]{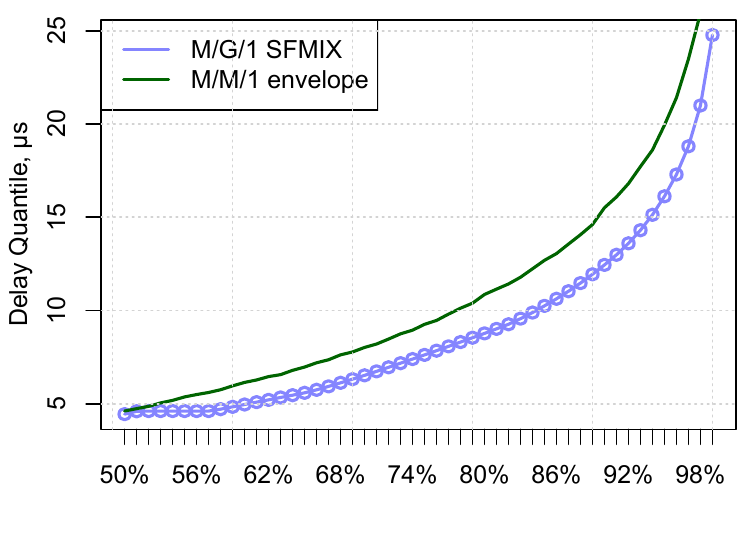}
    \caption{PDF of delay simulated packets (left) their percentiles (right) in one hop}
    \label{fig:PDF_quantiles}
\end{figure*}

Fig~\ref{fig:PDF_quantiles} (left) illustrates an example of the applicability of Algorithm~\ref{alg:envelope_calculation}. In the figure, the blue histogram are obtained from delays of an M/G/1 queue fed with real traffic following the SFM-IXP packet size distribution; in green, the M/M/1 queuing envelope, that is, an exponential distribution with mean $E(D_{\text{env}}) = E(X)\frac{1}{1-\rho_{\text{env}}}$. The Quantile function is also depicted in Fig~\ref{fig:PDF_quantiles} (right) where, as shown, the M/M/1 envelope model is always above the real one for percentiles above the median, acting as a tight upper delay bound. The scenario considers a queue fed with real SFM-IXP traffic (packet size distribution) operating at 10 Gbit/s and real load $\rho_{\text{real}}=0.7$. The algorithm finds that the smallest envelope load for using an M/M/1 as the upper bound is $\rho_{\text{env}}=0.78$. The average (real) delay is:
$$E(D_{\text{real}}) = 5.34~\mu s$$
while the average delay according to the envelope follows:
$$E(D_{\text{env}}) = E(X)\frac{1}{1-\rho_{\text{env}}} = 6.36~\mu s$$

The M/M/1 envelope also provides an exact formula (see eq.~\ref{eq:quantile}) for computing delay percentiles:
$$D_{0.90, \text{env}} = 1.40 \mu s \frac{1}{1-0.78}\ln \frac{1}{1-0.90} = 14.65 \mu s$$
$$D_{0.99, \text{env}} =  1.40 \mu s \frac{1}{1-0.78}\ln \frac{1}{1-0.99} = 29.31 \mu s$$
while the real quantile values, according to the histogram of Fig.~\ref{fig:PDF_quantiles} are $12.44~\mu s$ and $24.77~\mu s$, thus acting as slightly upper delay bounds.

\section{Simulations}
\label{Simulation}

To validate the M/M/1 envelope calculator defined in Algorithm~\ref{alg:envelope_calculation}, we use\texttt{simmer}~\cite{simmer}, a discrete-event simulator (DES) implemented in R, very useful to model queuing networks. 

Our methodology was applied to different packet size distributions commonly observed in real-world deterministic network environments, namely: 

\begin{enumerate}
    \item \textbf{Trimodal} A tri-modal distribution with packet sizes of 40 Bytes, 576 Bytes, and 1500 Bytes, with weighted probabilities of $\frac{7}{12}$, $\frac{4}{12}$, and $\frac{1}{12}$, respectively, found in~\cite{simmer}. This distribution has an average packet size of 340.33 bytes, a standard deviation of 428.01 bytes, and a coefficient of variation of $C_X^2 = 1.58$.

    \item \textbf{Amsterdam} A distribution based on traffic data from the AMS-IX Amsterdam Internet Exchange, which has an average packet size of 1019.03 Bytes, a standard deviation of 1161.66 Bytes.

    \item \textbf{San Francisco} A distribution based on traffic data from the SFM-IX San Francisco Metropolitan Internet Exchange, with parameters of 1750.41 Bytes for the average packet size, a standard deviation of 2062.69 Byte. 
    
 \end{enumerate}
    
By considering these three diverse service time distributions, the study aims to evaluate the accuracy and applicability of the M/M/1 envelope approximation across a range of traffic scenarios, from synthetic tri-modal distributions to real-world traffic patterns observed at different Internet exchange points. This approach allows for a comprehensive assessment of the approximation's performance and its suitability for modeling queuing delays in DetNet environments with varying packet size distributions and traffic characteristics.

\subsection{Characterisation of M/M/1 envelope load}

\begin{figure}[!t]
    \centering
    \includegraphics[width=0.9\columnwidth]{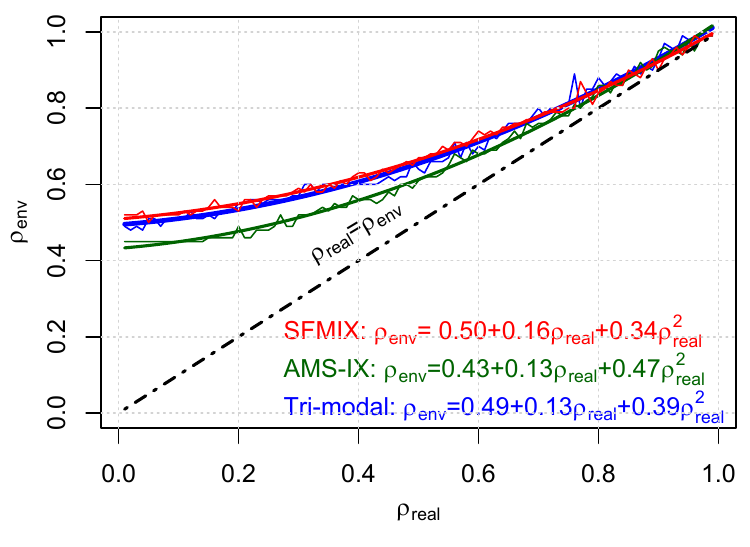}
    \caption{Relationship between M/G/1 real load and M/M/1 envelope bound}
    \label{fig:load_polynomial_regressions}
\end{figure}

Fig.~\ref{fig:load_polynomial_regressions} illustrates the relationship between the load of the M/M/1 envelope and the load of the M/G/1 models for different packet size distributions, along with fitted polynomial approximations. 


The equations for the fitted polynomials in Fig.~\ref{fig:load_polynomial_regressions} are:
\begin{itemize}
    \item Tri-modal: $\rho_{env} = 0.49 + 0.13\cdot\rho_{real} + 0.39\cdot \rho_{real}^2$
    \item AMS-IX: $\rho_{env} = 0.43 + 0.13\cdot\rho_{real} + 0.47\cdot \rho_{real}^2$
    \item SFM-IX: $\rho_{env} = 0.50 + 0.16\cdot\rho_{real} + 0.34\cdot \rho_{real}^2$
\end{itemize}

The worst-case scenario shown in Fig.~\ref{fig:load_polynomial_regressions} corresponds to the SFM-IX size distribution, which can be taken as a reference for the next section's numerical examples. 


Finally, it is worth noticing from the figures that as the system's load increases, the queueing delay distribution in the M/G/1 system gradually converges towards an exponential distribution with the mean of eq.~\ref{eq:mg1}, as stated by the Kingman's law of congestion~\cite{kingman_1961}. 
Consequently, the delay characteristics of both real and envelope models become increasingly similar as the system approaches saturation. This convergence suggests that the M/M/1 envelope provides a tighter upper bound approximation for the delays in the M/G/1 system under higher load conditions.


\subsection{Using the M/M/1 envelope model for DetNets}
As a \textbf{numerical example 1}, let us consider a scenario where a single node collects the traffic of $N=3,000$ households. The average traffic per user is approximately $\mu = 1~Mb/s$ with a standard deviation of $\sigma = 0.8~Mb/s$, in line with forecasted estimates for residential traffic of~\cite{ftth_hernandez}. We are interested in estimating the 90-th delay percentile of latency perceived by a packet chosen at random assuming the aggregated traffic traverses a node operating at 10~Gb/s.

Following the central limit theorem, we can assume that the aggregated traffic is normally distributed $N(N\mu, N\sigma^2)$, that is, with a mean of 3~Gb/s and standard deviation of 0.044~Gb/s. Our rule of dimensioning would be to consider as peak traffic $A_{\text{peak}} = 3+3 \cdot 0.044 = 3.13~Gb/s$. Thus, the system's load is:
$$\rho_{\text{real}} = \frac{3.13~Gb/s}{10~Gb/s} = 0.313$$
or 31.3\%. 

As previously stated, conducting real traffic simulations in this scenario may be computationally expensive. Thanks to the M/M/1 envelope, we can approximate the total delay that packets would suffer by first estimating the envelope's load:
$$\rho_{env} = 0.50+0.16\cdot 0.313 + 0.34\cdot 0.313^2 = 0.583$$
and then the average M/M/1 delay:
$$E(D_{env}) = E(X)\frac{1}{1-\rho_{env}} = 1.4\mu s \frac{1}{1-0.583}=3.35~\mu s$$
where $E(X)=\mu^{-1}=1.4\mu s$ for the average SFM-IX packet transmitted at 10~Gb/s.

Similarly, the 90-th and 99-th envelope percentiles are:
$$D_{0.90,env} = E(X)\frac{1}{1-\rho_{env}}\ln\frac{1}{1-0.9} = 7.71~\mu s$$
$$D_{0.99,env} = E(X)\frac{1}{1-\rho_{env}}\ln\frac{1}{1-0.99} = 15.43~\mu s$$

The \textbf{numerical example 2} illustrates the accuracy of the M/M/1 envelope approximation for a different network capacity of $C = 400~Gb/s$ and two distinct traffic distributions: Tri-modal and AMS-IX. The goal is to compare the simulated packet delays obtained from simulation with the delays computed using the derived M/M/1 envelope model. This comparison is performed by evaluating the average delays, as well as the 99th percentile of the delay distributions, against the corresponding quantiles of the simulated data.

\begin{figure}[!htbp]
    \centering
    \includegraphics[width=0.9\columnwidth]{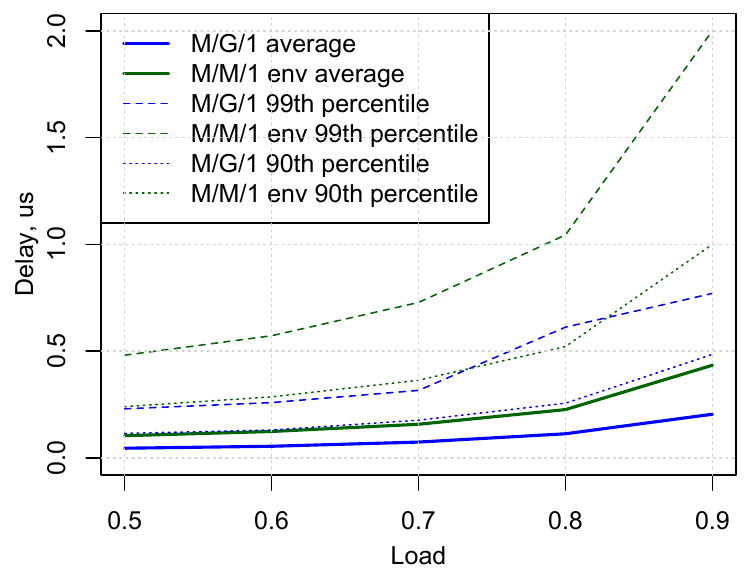}
    \caption{Comparison of theoretical M/M/1 envelope and simulated M/G/1 delays for AMS-IX traffic distribution}
    \label{fig:numeric_example_2_v2}
\end{figure}

Fig.~\ref{fig:numeric_example_2_v2} shows both real and M/M/1 envelope upper bound values for the average delay, 90-th and 99-th percentiles for various link loads at 400~Gb/s. 


The results demonstrate that the simulated delays are consistently below the corresponding theoretical values computed using the M/M/1 envelope model. This validates that the proposed M/M/1 envelope serves as a tight upper bound for the queuing delays experienced by IP packets in the Internet.

\section{Summary and discussion}
\label{conclusion}

This article has proposed a simple model to approximate real IP packet delay with and M/M/1 queuing delay envelope. To use the envelope model, the network designer only has to estimate the real traffic load of its router $\rho_{real}$, convert it to $\rho_{env}$ using the models of Fig.~\ref{fig:load_polynomial_regressions} and apply the well-known closed-form equations of M/M/1 queueing systems (eq.~\ref{eq:mm1_cdf}). This methodology allows to find delay percentile estimates for dimensioning DetNet scenarios where tight delay guarantees needs to be fulfilled.



The algorithm for finding M/M/1 envelopes and code for re-running this work's simulations can be found open-source in Github\cite{github_code}.

\section*{Acknowledgment}
The authors would like to acknowledge the support of Spanish projects ITACA (PDC2022-133888-I00) and EU project SEASON (grant no. 101096120).




\footnotesize

\bibliographystyle{IEEEtran}


\bibliography{references}

\end{document}